\documentclass{ws-procs9x6}

\newcommand{\be}{\begin{eqnarray}}
\newcommand{\ee}{\end{eqnarray}}
\newcommand{\non}{\nonumber}
\newcommand{\sh}{\mathop{\rm sh}\nolimits}
\newcommand{\ch}{\mathop{\rm ch}\nolimits}
\newcommand{\tnh}{\mathop{\rm th}\nolimits}
\newcommand{\cth}{\mathop{\rm cth}\nolimits}
\newcommand{\csch}{\mathop{\rm csch}\nolimits}
\newcommand{\sech}{\mathop{\rm sech}\nolimits}
\newcommand{\tr}{\mathop{\rm tr}\nolimits}
\newcommand{\id}{\mathbb{I}}

\begin{document}

\title{Bethe Ansatz for the open XXZ chain\\
from functional relations at roots of unity}

\author{Rafael I. Nepomechie\footnote{\uppercase{W}ork supported in
part by the \uppercase{N}ational \uppercase{S}cience
\uppercase{F}oundation under \uppercase{G}rant
\uppercase{PHY}-0244261.}}

\address{Physics Department, P.O. Box 248046, University of Miami\\
Coral Gables, FL 33124, USA\\
E-mail: nepomechie@physics.miami.edu}

\maketitle

\abstracts{
We briefly review Bethe Ansatz solutions of the integrable open spin-${1\over 2}$
XXZ quantum spin chain derived from functional relations obeyed by the 
transfer matrix at roots of unity.}

\section{Introduction}\label{sec:intro}

A long standing problem has been to solve the open spin-${1\over 2}$ XXZ
quantum spin chain with general integrable boundary terms, defined by
the Hamiltonian \cite{dVGR,GZ}
\be
{\mathcal H}&=& {1\over 2}\Big\{ \sum_{n=1}^{N-1}\left( 
\sigma_{n}^{x}\sigma_{n+1}^{x}+\sigma_{n}^{y}\sigma_{n+1}^{y}
+\ch \eta\ \sigma_{n}^{z}\sigma_{n+1}^{z}\right)\label{Hamiltonian} \\
&+& \sh \eta \Big[ 
\cth \alpha_{-} \tnh \beta_{-}\sigma_{1}^{z}
+ \csch \alpha_{-} \sech \beta_{-}\big( 
\ch \theta_{-}\sigma_{1}^{x} 
+ i\sh \theta_{-}\sigma_{1}^{y} \big) \non \\
& & \quad -\cth \alpha_{+} \tnh \beta_{+} \sigma_{N}^{z}
+ \csch \alpha_{+} \sech \beta_{+}\big( 
\ch \theta_{+}\sigma_{N}^{x}
+ i\sh \theta_{+}\sigma_{N}^{y} \big)
\Big] \Big\} \,, \non 
\ee
where $\sigma^{x} \,, \sigma^{y} \,, \sigma^{z}$ are the standard
Pauli matrices, $\eta$ is the bulk anisotropy parameter, $\alpha_{\pm} \,,
\beta_{\pm} \,, \theta_{\pm}$ are arbitrary boundary parameters, 
and $N$ is the number of spins.  
Determining the energy eigenvalues in terms of solutions of a system
of Bethe Ansatz equations is a fundamental problem, which has
important applications in integrable quantum field theory as well as
condensed matter physics and statistical mechanics \cite{Bat}, and
perhaps also string theory. (For an introduction to Bethe Ansatz, see 
e.g. Refs.~\refcite{Ba,KBI,Fa}.)

The basic difficulty in solving (\ref{Hamiltonian}) is that, 
in contrast to the special case of
diagonal boundary terms ({\it i.e.}, $\alpha_{\pm}$ or $\beta_{\pm} 
\rightarrow \pm \infty$, 
in which case ${\mathcal H}$ has a $U(1)$ symmetry) which was solved long 
ago \cite{Ga,ABBBQ,Sk}, a simple
pseudovacuum state does {\it not} exist. For instance, the state with all
spins up is not an eigenstate of the Hamiltonian.  Hence, many of the
techniques which have been developed to solve integrable models cannot
be applied.  

We observed some time ago \cite{XX,XXZ} that, for bulk anisotropy
parameter values
\be
\eta = {i \pi\over p+1}\,, \qquad p= 1 \,, 2 \,, \ldots 
\label{eta}
\ee
(hence $q \equiv e^{\eta}$ is a root of unity, satisfying
$q^{p+1}=-1$) and arbitrary values of the boundary parameters, 
the model's transfer matrix $t(u)$ (see Sec. 2) obeys a functional
relation of order $p+1$. For example, for the case $p=2$, 
the functional relation is
\be
t(u) t(u+\eta) t(u+2\eta) 
&-& \delta(u-\eta) t(u+\eta) 
- \delta(u) t(u+2\eta) \non \\
&-& \delta(u+\eta) t(u) 
= f(u)  \,,
\label{funcrltn}
\ee
where $\delta(u)$ and $f(u)$ are known scalar functions which depend
on the boundary parameters. (Expressions
for these functions in terms of the boundary parameters in
(\ref{Hamiltonian}) are given in Ref.~\refcite{special}.) Similar
results had been known earlier for closed spin chains.\cite{Ba2,BLZ,KSS}

By exploiting these functional relations, we have obtained Bethe
Ansatz solutions of the model for various special cases of the bulk
and boundary parameters:

\begin{itemize}
    
\item[(i)] [Refs.~\refcite{funcrltns,constraint,NR}]
The bulk anisotropy parameter has values (\ref{eta});
and the boundary parameters satisfy the constraint
\be 
\alpha_{-} + \beta_{-} + \alpha_{+} + \beta_{+} = \pm
(\theta_{-} - \theta_{+}) + \eta k \,,
\label{constraint}
\ee 
where $k \in [-(N+1) \,, N+1 ]$ 
is even (odd) if $N$ is odd (even), respectively.
    
\item[(ii)] [Ref.~\refcite{special}]
The bulk anisotropy parameter has values (\ref{eta}) 
with $p$ {\it even}; and either 

  \begin{itemize}
      
      \item[(a)] at most one of the boundary parameters is nonzero, or
    
      \item[(b)] any two of the boundary parameters 
$\{ \alpha_{-}, \alpha_{+},\beta_{-}, \beta_{+} \}$
are arbitrary, the remaining boundary parameters from this set
are either $\eta$ or $i \pi/2$, and $\theta_{-} = \theta_{+}$.

  \end{itemize} 
  
\item[(iii)]  [Ref.~\refcite{generalizedTQ}]
The bulk anisotropy parameter has values (\ref{eta}) 
with $p$ {\it odd}; at most two of the boundary parameters 
$\{ \alpha_{-}, \alpha_{+},\beta_{-}, \beta_{+} \}$
are nonzero, and $\theta_{-} = \theta_{+}$. 

\end{itemize} 

\noindent 
All of these cases have the property that the quantity $\Delta(u)$,
defined by
\be 
\Delta(u) =  f(u)^{2} -4  \prod_{j=0}^{p} \delta(u+j\eta) \,,
\label{Delta}
\ee 
is a perfect square.

Solutions for generic values of the bulk anisotropy parameter and 
for boundary parameters obeying a constraint similar to (\ref{constraint}) 
have been discussed in Refs.~\refcite{CLSW,dGP,YZG}.

Here we briefly review our results for the cases (i) - (iii).

\section{Transfer matrix}\label{sec:transfer}

The transfer matrix $t(u)$ of the open XXZ chain with general integrable
boundary terms, which satisfies the fundamental commutativity property
$\left[ t(u)\,, t(v) \right] = 0$, is given by \cite{Sk}
\be
t(u) = \tr_{0} K^{+}_{0}(u)\  
T_{0}(u)\  K^{-}_{0}(u)\ \hat T_{0}(u)\,,
\label{transfer}
\ee
where $T_{0}(u)$ and $\hat T_{0}(u)$ are the monodromy matrices 
\be
T_{0}(u) = R_{0N}(u) \cdots  R_{01}(u) \,,  \qquad 
\hat T_{0}(u) = R_{01}(u) \cdots  R_{0N}(u) \,,
\label{monodromy}
\ee
and $\tr_{0}$ denotes trace over the ``auxiliary space'' 0.
The $R$ matrix is given by
\be
R(u) = \left( \begin{array}{cccc}
	\sh  (u + \eta) &0            &0           &0            \\
	0                 &\sh  u     &\sh \eta  &0            \\
	0                 &\sh \eta   &\sh  u    &0            \\
	0                 &0            &0           &\sh  (u + \eta)
\end{array} \right) \,,
\label{bulkRmatrix}
\ee 
where $\eta$ is the bulk anisotropy parameter; and $K^{\mp}(u)$ are
$2 \times 2$ matrices whose components
are given by \cite{dVGR,GZ}
\be
K_{11}^{-}(u) &=& 2 \left( \sh \alpha_{-} \ch \beta_{-} \ch u +
\ch \alpha_{-} \sh \beta_{-} \sh u \right) \non \\
K_{22}^{-}(u) &=& 2 \left( \sh \alpha_{-} \ch \beta_{-} \ch u -
\ch \alpha_{-} \sh \beta_{-} \sh u \right) \non \\
K_{12}^{-}(u) &=& e^{\theta_{-}} \sh  2u \,, \qquad 
K_{21}^{-}(u) = e^{-\theta_{-}} \sh  2u \,,
\label{Kminuscomponents}
\ee
and
\be
K^{+}(u) = K^{-}(-u-\eta)
 \Bigg\vert_{ 
 \alpha_{-}\rightarrow -\alpha_{+}\,, \ 
 \beta_{-}\rightarrow -\beta_{+}\,, \
 \theta_{-}\rightarrow \theta_{+}} \,,
\ee
where $\alpha_{\mp} \,, \beta_{\mp} \,, \theta_{\mp}$ are the boundary
parameters. The Hamiltonian (\ref{Hamiltonian}) is proportional
to $t'(0)$ plus a constant.

The transfer matrix also has $i \pi$ periodicity
\be
t(u+ i \pi) = t(u) \,,
\label{periodicity}
\ee
crossing symmetry
\be
t(-u - \eta)= t(u) \,,
\label{transfercrossing}
\ee
and the asymptotic behavior 
\be
t(u) \sim -\ch(\theta_{-}-\theta_{+})
{e^{u(2N+4)+\eta (N+2)}\over 2^{2N+1}} \id + 
\ldots \qquad \mbox{for} \qquad
u\rightarrow \infty \,.
\label{transfasympt}
\ee

\section{Case (i)}\label{sec:case(i)}

Our main objective is to determine the eigenvalues $\Lambda(u)$
of the transfer matrix $t(u)$ (\ref{transfer}), from which the
energy eigenvalues can readily be computed.
The functional relations for the transfer matrix (e.g.,
(\ref{funcrltn})) evidently
imply corresponding relations for $\Lambda(u)$.  
Following Ref.~\refcite{BR}, we observe that the latter 
relations can be written as
\be
\det {\mathcal M}(u) = 0 \,,
\label{detzero}
\ee
where ${\mathcal M}(u)$ is the $(p+1) \times (p+1)$ matrix
\be
{\mathcal M}(u) = \left(
\begin{array}{cccccccc}
    \Lambda(u) & -{\delta(u)\over h(u+\eta)} & 0  & \ldots  & 0 & -h(u)  \\
    -h(u+\eta) & \Lambda(u+\eta) & -{\delta(u+\eta)\over h(u+2\eta)} & \ldots  & 0 & 0  \\
    \vdots  & \vdots & \vdots & \ddots 
    & \vdots  & \vdots    \\
   -{\delta(u-\eta)\over h(u)}  & 0 & 0 & \ldots  & -h(u+p\eta) &
    \Lambda(u+p\eta)
\end{array} \right) 
\label{calM}
\ee
if there exists an $i \pi$-periodic function $h(u)$ such that 
\be
f(u) = \prod_{j=0}^{p} h(u+j\eta)
+ \prod_{j=0}^{p}{\delta(u+j\eta) \over h(u+j\eta)} \,.
\label{tough}
\ee
To solve for $h(u)$, we set $z(u) \equiv \prod_{j=0}^{p}
h(u+j\eta)$, and observe that (\ref{tough}) implies that $z(u)$
satisfies a quadratic equation
\be
z(u)^{2} -z(u) f(u) +  \prod_{j=0}^{p} \delta(u+j\eta) = 0 \,,
\label{quadratic}
\ee
whose solution is evidently given by
\be 
z(u) = {1\over 2}\left( f(u) \pm \sqrt{\Delta(u)} \right)  \,,
\label{zsltn}
\ee
where $\Delta(u)$ is defined in (\ref{Delta}).
If the boundary parameters satisfy the constraint (\ref{constraint}), 
then $\Delta(u)$ is a perfect square, and two solutions of (\ref{tough}) are  
\be
h^{(\pm)}(u) &=& -4 \sh^{2N}(u+\eta){\sh(2u+2\eta)\over \sh(2u+\eta)} 
\non \\
&\times& \sh(u \pm \alpha_{-}) \ch(u \pm \beta_{-}) 
\sh(u \pm \alpha_{+}) \ch(u \pm \beta_{+}) \,.
\label{hpm}
\ee
Let us now label the corresponding matrices (\ref{calM}) by ${\mathcal
M}^{(\pm)}(u)$.

The condition (\ref{detzero}) implies that 
${\mathcal M}^{(\pm)}(u)$ has a null eigenvector $v^{(\pm)}(u)$,
\be
{\mathcal M}^{(\pm)}(u)\ v^{(\pm)}(u) = 0 \,, 
\label{nulleigenvec}
\ee
Note that the matrix ${\mathcal M}^{(\pm)}(u)$ has the symmetry
\be
S {\mathcal M}^{(\pm)}(u) S^{-1} = {\mathcal M}^{(\pm)}(u+\eta) \,,
\label{calMsymmetry}
\ee
where 
\be
S = \left(
\begin{array}{cccccccc}
    0 & 1 & 0  & \ldots  & 0 & 0  \\
    0 & 0 & 1  & \ldots  & 0 & 0  \\
    \vdots  & \vdots & \vdots & \ddots 
    & \vdots  & \vdots    \\
    0 & 0 & 0  & \ldots  & 0 & 1 \\
   1  & 0 & 0 & \ldots  & 0 & 0
\end{array} \right) \,, \qquad S^{p+1} =  \id \,.
\label{Smatrix}
\ee
It follows that the null eigenvector $v^{(\pm)}(u)$ satisfies 
$S v^{(\pm)}(u) = v^{(\pm)}(u + \eta)$. Thus, its components can be expressed in terms of a
function $Q^{(\pm)}(u)$,
\be
v^{(\pm)}(u) = \big( Q^{(\pm)}(u)\,, Q^{(\pm)}(u+\eta) \,, \ldots \,, 
Q^{(\pm)}(u+p\eta) \big) \,,
\label{vexplicit}
\ee
with $Q^{(\pm)}(u + i\pi) = Q^{(\pm)}(u)$. We make the Ansatz
\be
Q^{(\pm)}(u) = \prod_{j=1}^{M^{(\pm)}} 
\sh(u - u_{j}^{(\pm)}) \sh(u + u_{j}^{(\pm)} + \eta) \,,
\label{Qpm}
\ee
which has the crossing symmetry $Q^{(\pm)}(u) = Q^{(\pm)}(-u-\eta)$.
Substituting the expressions for ${\mathcal M}^{(\pm)}(u)$ (\ref{calM}) and 
$v^{(\pm)}(u)$ (\ref{vexplicit}) into the null eigenvector equation 
(\ref{nulleigenvec}) yields the result for the transfer matrix
eigenvalues
\be
\Lambda^{(\pm)}(u) = h^{(\pm)}(u) {Q^{(\pm)}(u-\eta)\over Q^{(\pm)}(u)} 
+ h^{(\pm)}(-u-\eta) {Q^{(\pm)}(u+\eta)\over Q^{(\pm)}(u)}  \,.
\label{Lambda}
\ee
The asymptotic behavior (\ref{transfasympt}) implies that 
$M^{(\pm)}={1\over 2}(N-1\pm k)$, where $k$ is the integer appearing
in the constraint (\ref{constraint}). Analyticity of the eigenvalues (\ref{Lambda})
implies the Bethe Ansatz equations
\be
{h^{(\pm)}(u_{j}^{(\pm)})\over h^{(\pm)}(-u_{j}^{(\pm)}-\eta)} = 
-{Q^{(\pm)}(u_{j}^{(\pm)}+\eta)\over Q^{(\pm)}(u_{j}^{(\pm)}-\eta)} \,, 
\qquad j = 1 \,, \ldots \,, M^{(\pm)} \,.
\label{BAE}
\ee
In short, for case (i), the eigenvalues of the transfer matrix (\ref{transfer})
are given by (\ref{Lambda}), where $h^{(\pm)}(u)$ and $Q^{(\pm)}(u)$
are given by (\ref{hpm}), (\ref{Qpm}) and (\ref{BAE}).

In Ref.~\refcite{NR}, we have verified numerically that this solution holds also
for generic values of $\eta$, which is consistent with
Refs.~\refcite{CLSW,dGP,YZG}; and that this solution gives the complete set of 
$2^{N}$ eigenvalues. To illustrate how completeness is achieved, let
us consider the case $N=4$. The integer $k$ in the constraint  
(\ref{constraint}) must therefore be odd, with $-5 \le k \le 5$. The
six possibilities are summarized in Table 1.
\begin{table}[ph]
    \tbl{Completeness for $N=4$. For each $k$, there are
    $2^{4}$ eigenvalues.}
{\footnotesize
\begin{tabular}{r|c|c}
    \hline
   $k $ & \# eigenvalues given by $\Lambda^{(+)}(u)$  & 
\# eigenvalues given by $\Lambda^{(-)}(u)$ \\
   \hline
   5 & 16 & 0 \\
   3 & 15 & 1 \\
   1 & 11 & 5 \\
   -1 & 5 & 11 \\
   -3 & 1 & 15 \\
   -5 & 0 & 16 \\ 
   \hline
   \end{tabular} }
   \vspace*{-13pt}
   \end{table}

\section{Case(ii)}

A key feature of case (i) is that the quantity $\Delta(u)$ 
(\ref{Delta}) is a perfect square. We therefore look for additional
such cases. For $p$ even, we find that
$\Delta(u)$ is also a perfect square if either (a) at most 
one of the boundary parameters is nonzero; or (b) any two of the boundary 
parameters $\{ \alpha_{-}, \alpha_{+},\beta_{-}, \beta_{+} \}$
are arbitrary, the remaining boundary parameters from this set
are either $\eta$ or $i \pi/2$, and $\theta_{-} = \theta_{+}$.
For definiteness, we focus here on the subcase (b) with $\alpha_{\pm}$
arbitrary, $\beta_{\pm}=\eta$
and $N$ even.
Unfortunately, the resulting $z(u)$ (\ref{zsltn}) is not consistent. 
To surmount this difficulty, we use a matrix ${\mathcal M}(u)$ 
which is different from (\ref{calM}), namely \cite{special}
\be
\lefteqn{{\mathcal M}(u)=}  \\ 
&&\left(
\begin{array}{cccccccc}
    \Lambda(u) & -h(u) & 0  & \ldots  & 0 & -h(-u+p \eta)  \\
    -h(-u) & \Lambda(u+p\eta) & -h(u+p \eta)  & \ldots  & 0 & 0  \\
    \vdots  & \vdots & \vdots & \ddots 
    & \vdots  & \vdots    \\
   -h(u+p^{2} \eta)  & 0 & 0 & \ldots  & -h(-u-p(p-1) \eta) &
    \Lambda(u+p^{2}\eta)
\end{array} \right)  \non 
\ee
where $h(u)$ is $2i\pi$-periodic. This matrix has the symmetry
\be
S {\mathcal M}(u) S^{-1} = {\mathcal M}(u+p\eta) \,,
\ee
where $S$ is given by (\ref{Smatrix}). By arguments similar to those
used in Sec. 3, we find that the transfer matrix eigenvalues 
are given by 
\be
\Lambda(u) = h(u) {Q(u + p\eta)\over Q(u)} 
+ h(-u+p \eta) {Q(u -p\eta)\over Q(u)}  \,,
\ee
where $h(u)$ is given by 
\be
h(u) &=&  4\sh^{2N}(u+\eta){\sh(2u+2\eta)\over \sh(2u+\eta)}
\ch^{2}(u -\eta)    \\
&\times& \sh(u-\alpha_{-}) \sh(u+\alpha_{+})
{\ch\left({1\over 2}(u+\alpha_{-}+\eta) \right)\over 
 \ch\left({1\over 2}(u-\alpha_{-}-\eta) \right)} 
 {\ch\left({1\over 2}(u-\alpha_{+}+\eta) \right)\over 
 \ch\left({1\over 2}(u+\alpha_{+}-\eta) \right)}\,, \non
\ee
and $Q(u)$ is given by
\be
Q(u) = \prod_{j=1}^{M} 
\sh \left( {1\over 2}(u - u_{j}) \right)
\sh \left( {1\over 2}(u + u_{j} - p\eta) \right) \,,
\ee 
with $M=N+2p+1$; and the Bethe Ansatz equations are
\be
{h(u_{j})\over h(-u_{j}+p\eta)} = 
-{Q(u_{j}-p\eta)\over Q(u_{j}+p\eta)} \,, 
\qquad j = 1 \,, \ldots \,, M \,.
\ee
We have verified numerically the completeness of this solution. The
other subcases (a) and (b) are mostly similar.  \footnote{The exception is 
the subcase (a) with $\theta_{\pm}$ nonzero, for which 
$Q(u) = \prod_{j=1}^{2M} \sh (u - u_{j})$, which is not
crossing symmetric. See Sec. 3.3 in Ref.~\refcite{special}.}
 
\section{Case(iii)}

For $p$ odd, we find that the quantity
$\Delta(u)$ (\ref{Delta}) is also a perfect square if at most two of the boundary parameters 
$\{ \alpha_{-}, \alpha_{+},\beta_{-}, \beta_{+} \}$
are nonzero, and $\theta_{-} = \theta_{+}$. 
For definiteness, we focus here on the case with $\alpha_{\pm}$
arbitrary, $\beta_{\pm}=0$ and $N$ even.
As in case (ii), the resulting 
$z(u)$ (\ref{zsltn}) is not consistent. To surmount this difficulty, we 
again use a matrix ${\mathcal M}(u)$ 
which is different from (\ref{calM}), namely \cite{generalizedTQ}
\be
\lefteqn{{\mathcal M}(u)=} \label{newM} \\ 
&&\left(
\begin{array}{cccccccc}
    \Lambda(u) & -{\delta(u)\over h^{(1)}(u)} & 0  & \ldots  & 0 &
    -{\delta(u-\eta)\over h^{(2)}(u-\eta)}  \\
    -h^{(1)}(u) & \Lambda(u+\eta) & -h^{(2)}(u+\eta)  & \ldots  & 0 & 0  \\
    \vdots  & \vdots & \vdots & \ddots 
    & \vdots  & \vdots    \\
   -h^{(2)}(u-\eta)  & 0 & 0 & \ldots  & -h^{(1)}(u+(p-1)\eta) &
    \Lambda(u+p\eta) 
\end{array} \right)   \non 
\ee
where $h^{(1)}(u)$ and $h^{(2)}(u)$ are $i\pi$-periodic.
It has the reduced symmetry
\be
T {\mathcal M}(u) T^{-1} = {\mathcal M}(u+2\eta) \,,
\label{calMreducedsymmetry}
\ee
where $T=S^{2}$, and $S$ is given by (\ref{Smatrix}). (While (\ref{calMsymmetry}) 
implies (\ref{calMreducedsymmetry}), the converse is not true.)
The condition $\det {\mathcal M}(u) =
0$ implies that ${\mathcal M}(u)$ has a null eigenvector $v(u)$,
\be
{\mathcal M}(u)\ v(u) = 0 \,, 
\label{nulleigenveciii}
\ee
where $v(u)$ satisfies $T v(u) = v(u + 2\eta)$. Thus, its components
are expressed in terms of {\it two} independent functions $Q_{1}(u)$,
$Q_{2}(u)$:
\be
v(u) = \left( Q_{1}(u)\,, Q_{2}(u) \,, \ldots \,, Q_{1}(u-2\eta) 
\,, Q_{2}(u-2\eta)\right) \,.
\label{vexplicitiii}
\ee
We make the Ans\"atze 
\be
Q_{1}(u) &=& \prod_{j=1}^{M_{1}} 
\sh (u - u_{j}^{(1)}) \sh (u + u_{j}^{(1)} + \eta) \,, \non \\
Q_{2}(u) &=& \prod_{j=1}^{M_{2}} 
\sh (u - u_{j}^{(2)}) \sh (u + u_{j}^{(2)} + 3\eta) \,.
\ee 
Substituting the expressions for ${\mathcal M}(u)$ (\ref{newM}) and 
$v(u)$ (\ref{vexplicitiii}) into the null eigenvector equation 
(\ref{nulleigenveciii}) yields {\it two} expressions for the transfer matrix
eigenvalues, 
\be
\Lambda(u) &=& 
{\delta(u)\over h^{(1)}(u)} {Q_{2}(u)\over Q_{1}(u)} 
+ {\delta(u-\eta)\over h^{(2)}(u-\eta)} {Q_{2}(u-2\eta)\over 
Q_{1}(u)} \,, \non \\
 &=& 
h^{(1)}(u-\eta) {Q_{1}(u-\eta)\over Q_{2}(u-\eta)} 
+ h^{(2)}(u) {Q_{1}(u+\eta)\over 
Q_{2}(u-\eta)} \,.
\label{genTQ}
\ee
Analyticity of these expressions leads to the Bethe Ansatz equations 
\be
{\delta(u_{j}^{(1)})\ h^{(2)}(u_{j}^{(1)}-\eta)
\over \delta(u_{j}^{(1)}-\eta)\ h^{(1)}(u_{j}^{(1)})} 
&=&-{Q_{2}(u_{j}^{(1)}-2\eta)\over Q_{2}(u_{j}^{(1)})} \,, \qquad j =
1\,, 2\,, \ldots \,, M_{1} \,, \non \\
{h^{(1)}(u_{j}^{(2)})\over h^{(2)}(u_{j}^{(2)}+\eta)}
&=&-{Q_{1}(u_{j}^{(2)}+2\eta)\over Q_{1}(u_{j}^{(2)})} \,, \qquad j =
1\,, 2\,, \ldots \,, M_{2} \,.
\ee
We expect that there are sufficiently many equations to determine all
the zeros $\{ u_{j}^{(1)} \,, u_{j}^{(2)} \}$ of $Q_{1}(u) \,, Q_{2}(u)$, 
respectively. 
Functions $h^{(1)}(u)$ (with $h^{(2)}(u) = h^{(1)}(-u-2\eta)$) which
ensure the condition $\det {\mathcal M}(u) = 0$ are given by
\be
h^{(1)}(u) &=& 4 \sh^{2N}(u + 2\eta) \,, \quad 
M_{2} = {1\over 2} N + {1\over 2}(3p - 1) \,, \quad M_{1} = M_{2} + 2  \,,
\non \\
p &=& 3\,, 7\,, 11\,, \ldots
\label{h11app}
\ee
and
\be
h^{(1)}(u) = \left\{ 
\begin{array}{ll}
    -2\ch(2u) \sh^{2}u \sh^{2N}(u + 2\eta)\,, 
    & M_{1} =M_{2} = {1\over 2} N + 2p - 1 \,, \\
\qquad \qquad p= 9\,, 17\,, 25\,, \ldots \\
   2\ch(2u) \sh^{2}u \sh^{2N}(u + 2\eta)\,,  
    & M_{1} =M_{2} = {1\over 2} N + {3\over 2}(p - 1) \,, \\
\qquad \qquad p= 5\,, 13\,, 21\,, \ldots \\
   2\ch(2u) \sh^{2}u  \sh^{2N}(u +
    2\eta)\,, & M_{1} =M_{2} = {1\over 2} N + 2 \,, \\
\qquad \qquad p=1 \,.
\end{array} \right.
\label{h12app}
\ee
We have verified numerically the completeness of this solution.
Similar results hold for the case with $\alpha_{-}, \beta_{-}$
arbitrary and $\alpha_{+}=\beta_{+}=0$, {\it etc}.

We observe that this solution represents a generalization of the
famous Baxter $T-Q$ relation \cite{Ba}, which schematically
has the form
\be
t(u)\ Q(u) = Q(u') + Q(u'') \,.
\ee
Indeed, our result (\ref{genTQ}) has the structure 
\be
t(u)\ Q_{1}(u) &=& Q_{2}(u') + Q_{2}(u'') \,, \non \\
t(u)\ Q_{2}(u) &=& Q_{1}(u') + Q_{1}(u'') \,.
\ee
Such generalized $T-Q$ relations, involving two or more
independent $Q(u)$'s, may also appear in other integrable models.

\section{Conclusions}

We have seen that Bethe Ansatz solutions of the open spin-${1\over 2}$ XXZ
quantum spin chain are available for the cases (i)-(iii), for which
the quantity $\Delta(u)$ (\ref{Delta}) is a perfect square. There may be 
further special cases for which $\Delta(u)$ is a 
perfect square, in which case it should not be difficult to find the 
corresponding Bethe Ansatz solution. Our
solution for case (iii) involves more than one $Q(u)$. This is a novel
structure, which should be further understood. The general case that  $\Delta(u)$
is not a perfect square and/or that $\eta \ne i\pi/(p+1)$ also remains to
be understood.

\section*{Acknowledgments}
Some of the work described here was done in collaboration with R. Murgan
and F. Ravanini. 
I am grateful to the conference organizers for their wonderful hospitality
and for the opportunity to present this work. I am also grateful to many of the
participants, in particular F.C. Alcaraz, L. Faddeev, J.-M. Maillard, G. Sierra, 
F.Y. Wu and W.-L. Yang, for their questions or comments.

\end{document}